\documentclass{PoS}

\title{Computing Nucleon Electric Dipole Moment from lattice QCD}

\ShortTitle{Computing Nucleon Electric Dipole Moment from lattice QCD}

\author{
Taku Izubuchi${}^{a,b}$,
\speaker{Hiroshi Ohki}${}^{a,c}$, 
Sergey~Syritsyn${}^{a,d}$, \\
  \llap{${}^a$} RIKEN/BNL Research Center, Brookhaven National Laboratory, 
      Upton, NY 11973, USA\\
  \llap{${}^b$} Physics Department, Brookhaven National Laboratory, Upton, New York 11973, USA\\
  \llap{${}^c$} Department of Physics, Nara Women's University, Nara 630-8506, Japan \\
  \llap{${}^d$} Department of Physics and Astronomy, Stony Brook University, 
      Stony Brook, NY 11794, USA \\
        E-mail: \email{hohki@asuka.phys.nara-wu.ac.jp}}

\abstract{
Electric dipole moments (EDMs) of nucleons and 
nuclei are actively considered as direct evidence of the CP violation.
Calculations of nucleon EDMs on lattice 
are required to connect the quark- and hadron- level effective CP violating interactions 
within QCD or other CP violating sources in new physics beyond the standard model.
Among them, 
the theta-induced nucleon EDM, that is 
the only such renormalizable interaction, 
has widely been investigated on a lattice.
In the report, we review recent developments of the lattice calculations of 
nucleon EDM induced QCD theta term.  

}

\FullConference{37th International Symposium on Lattice Field Theory - Lattice2019\\
		16-22 June 2019\\
		Wuhan, China}

\usepackage{cancel}
\usepackage{amsmath}

\newcommand{\CP}{{CP}}

\newcommand{\CPviol}{{\cancel{\rm CP}}}

\newcommand{\Slash}[1]{{\ooalign{\hfil/\hfil\crcr$#1$}}}

\begin{document}

\section{Introduction}
Observation of a  non-zero nucleon electric dipole moment (nEDM) 
would be direct evidence for violation of CP(T)-symmetry.
The Standard Model (SM) has a CP-violating phase in the CKM matrix.
However, there is at least one substantial motivation to search for CP violation beyond the SM: 
the magnitude of the CP violation effect in CKM is insufficient to explain the observed baryon asymmetry of the universe 
as required by one of the famous Sakharov's conditions for the universe's matter origin.
Currently the most sensitive probes for the CP-violating phenomena are EDM searches in hadronic, atomic, and molecular systems. 
The best limits on EDMs come from experiments on neutrons (ILL) \cite{Baker:2006ts} 
and $^{199}$Hg \cite{Graner:2016ses} which constrain the nucleon EDM (nEDM) 
$|d_n| \leq 2.6 \times 10^{-26}$ [$e$ cm]. 
This bound is $\sim 10^5$ larger than the prediction $\sim 10^{-31}$ [$e $ cm] 
from the CP phase of the CKM matrix in the SM. 
Observation of non-zero nEDMs would be a discovery of fundamental importance, 
and however, even a null result would make serious impact on cosmology and high-energy theory. 
Violations of CP symmetry at the quark level are represented by a number of effective quark and gluon operators. 
Among them, the only such renormalizable interaction is the QCD "$\theta$-term", 
\begin{align}
iS_{\theta} = i\theta Q=i\theta \sum_{x}q(x)\,, \quad \quad 
q(x) = \frac{1}{16\pi^2} {\rm Tr} \left[G_{\mu\nu}\tilde{G}^{\mu\nu}(x)\right]
\label{eqn:Qtopo}
\end{align}
where $Q$ is the topological charge and 
$q(x)$ is the topological charge density operator.
Assuming that the $\theta$-term is the only source of the CP violation, 
we have a strong constraint on the parameter $|\theta| \lesssim 10^{-10}$.
The problem of why $\theta$ is so small is known as the 
strong CP problem~\footnote{A dynamical solution to the strong CP problem is proposed in a parallel talk by G. Schierholz~\cite{Nakamura:2019ind}.}.
The precision of EDM measurements of nucleons and nuclei will increase in the future experiments using neutron sources, 
which plan to improve neutron EDMs bounds by 1-2 orders of magnitude.
However, quantitative connection between magnitudes of EDMs and such CP violation operators at the quark-gluon level 
is very limited and model-dependent 
(see Refs.~\cite{Engel:2013lsa}, \cite{Yamanaka:2016umw} for a recent review of the EDM phenomenology).
Therefore connecting the quark- and hadron-level effective interactions that include CP violating sources is an urgent task for lattice QCD.

In this proceedings, we review recent progress on the lattice calculations of the nEDM induced by the $\theta$ term. 
As for other CP-violating matrix elements that arise from the operators beyond the SM, 
these are discussed in the plenary lecture \cite{Tanmoy:Lattice2020}~\footnote{See also \cite{Gupta:2019fex} for a recent review.}.

\section{CP odd Form factor and parity mixing}

In this section, we briefly review the methods for the lattice computations of the nEDM.
The $\theta$-induced nEDMs have been calculated on a lattice 
from the CP-odd electric dipole form factor (EDFF) $F_3(Q^2)$ with the $Q^2 \to0$  extrapolation 
of the nucleon matrix elements of the quark vector current \cite{Shintani:2005xg, Berruto:2005hg, Aoki:2008gv, Guo:2015tla, 
Alexandrou:2015spa, Shintani:2015vsx, Abramczyk:2017oxr,Dragos:2018uzd, Dragos:2019oxn,Yoon:2020soi}, 
and from nucleon energy shifts in a uniform background electric field \cite{Shintani:2006xr, Shintani:2008nt, Abramczyk:2017oxr}.
The EDFF is defined as 
\begin{equation}
\label{eqn:ff_cpviol}
\langle p^\prime,\sigma^\prime |J^\mu|p,\sigma
\rangle_{\CPviol}
  = \bar{u}_{p^\prime,\sigma^\prime} 
  \big[F_1(Q^2) \gamma^\mu + \big(F_2(Q^2) 
  + iF_3(Q^2)\big) \frac{i\sigma^{\mu\nu}q_\nu}{2M_N} \big] u_{p,\sigma}\,,
\end{equation}
where $Q^2=-q^2$ and $q=p^\prime-p$, 
and $F_1$ and $F_2$ are the Dirac and Pauli form factors.
The QCD $\theta$ term is introduced either as a modification of the lattice CP-even QCD action with 
imaginary $\theta$ term~\cite{Aoki:2008gv,Guo:2015tla} 
or as a Taylor expansion of nucleon correlation functions, 
$S_\text{QCD}\to S_\text{QCD} + i S_\theta$. 
In the latter case the nucleon-current correlation functions in $\CPviol$ QCD vacuum are 
modified as 
\begin{gather}
\label{eqn:corr_cpviol}
\begin{aligned}
\langle N\,[\bar q \gamma^\mu q]\, \bar N \rangle_{\CPviol}
&= \frac1Z\int\,{\mathcal{D}} U\,\mathcal{D}\bar\psi\mathcal{D}\psi 
N\,[\bar q \gamma^\mu q]\, \bar N e^{-S - iS_\theta}\,  \\
&= C_{NJ\bar N} - i \theta\, C_{NJ\bar N}^Q
    + O(\theta^2)\,,
\end{aligned}
\end{gather}
\\
where
$C_{NJ\bar N} = \langle N\,[\bar q \gamma^\mu q]\, \bar N\rangle$ and 
$C_{NJ\bar N}^Q = \langle N\,[\bar q \gamma^\mu q]\, \bar N \, 
\sum_x [q(x)] \rangle$
are the nucleon-current correlation functions evaluated in the $\CP$-even QCD vacuum. 
To obtain the EDFF $F_3(Q^2)$ in Eq.~\eqref{eqn:ff_cpviol} we have to calculate 
the nucleon two and three point functions 
\begin{align}
\label{eq:c2pt}
C_{\rm 2pt}^Q(\vec{p},t) &= \sum_{\vec y} e^{-i\vec{p}\cdot\vec{y}} \langle N(\vec{y},t) \bar{N}(\vec{0},0) 
\sum_x[q(x)] \rangle, \\ 
\label{eq:c3pt}
C_{\rm 3pt}^Q(t, \vec{p};t_{\rm op}, \vec{q}) &= \sum_{\vec y, \vec z} e^{-i\vec{p}\cdot \vec{y}+\vec{q}\cdot \vec{z}} 
\langle N(\vec{y},t) J^\mu(\vec{z},t_{\rm op})\bar{N}(\vec{0},0) \sum_x[q(x)] \rangle.
\end{align}
In general the nucleon ground states as well as their overlaps with the positive-parity nucleon 
ground state are modified in $\CPviol$ QCD vacuum as 
\begin{align}
\langle 0| N|p,\sigma \rangle_\CPviol = Z_N e^{i\alpha\gamma_5} u_{p,\sigma}=Z_N \tilde{u}_{p,\sigma},
\end{align}
where $\tilde{u}_{p,\sigma}$ is a spinor wave function for the nucleon state $|p,\sigma\rangle_\CPviol$ 
and $Z_N$ is a normalization constant. 
The spinor $\tilde{u}_{p,\sigma}$ satisfies the following free Dirac equation with CP-violating $\gamma_5$ mass
\begin{align}
(\Slash{p}-m_Ne^{-2i\alpha \gamma_5})\tilde{u}_{p,\sigma} = 
(\Slash{p}-m_Ne^{-2i\alpha \gamma_5}) e^{i\alpha\gamma_5}  u_{p,\sigma} = 0,
\end{align}
where $u_{p,\sigma}$ is a wave function spinor in CP-even QCD vacuum.
Thus when the CP violating effect exists in the QCD vacuum, 
one should use the modified nucleon spinor $\tilde{u}_{p,\sigma}$ in lattice calculations 
which affects the kinematic coefficients.
For example,  ignoring excite states,  
the nucleon two-point correlation function with CP violating operator can be represented as 
\begin{align}
C_{\rm 2pt}^Q(\vec{p},t)
&= 
|Z_N|^2 \frac{e^{-E_p t}}{2E_p} \sum_{\sigma} \tilde{u}_{p,\sigma} \bar{\tilde{u}}_{p, \sigma}  
=
|Z_N|^2 \frac{e^{-E_p t}}{2E_p}
\left( m_N e^{2i\alpha\gamma_5} -i\Slash{p} \right), 
\end{align}
where we use the completeness condition for the free Dirac spinor, 
\begin{align}
\label{eq:2ptcpviol}
\sum_\sigma \tilde{u}_{p,\sigma} \bar{\tilde{u}}_{p, \sigma}=
e^{i\alpha\gamma_5} 
\left( 
\sum_\sigma u_{p,\sigma} \bar{u}_{p, \sigma}
\right) =m_N e^{2i\alpha\gamma_5}-i\Slash{p}.
\end{align}
The modification also affects the nucleon matrix elements $\langle p^\prime,\sigma^\prime |J^\mu|p,\sigma\rangle_\CPviol$ as
\begin{align}
\bar{\tilde{u}}_{p',\sigma'}
\left[
\tilde{F_1}\gamma^\mu + (\tilde{F_2}+i\tilde{F_3}\gamma_5) \frac{i\sigma^{\mu\nu}q_\nu}{2m_N}
\right]
\tilde{u}_{p,\sigma}
&=
\bar{u}_{p',\sigma'}
\left[
\tilde{F_1}\gamma^\mu + 
e^{2i\alpha\gamma_5} (\tilde{F_2}+i\tilde{F_3}\gamma_5) \frac{i\sigma^{\mu\nu}q_\nu}{2m_N}
\right]
u_{p,\sigma}
\nonumber \\
&=
\bar{u}_{p',\sigma'}
\left[
F_1\gamma^\mu + 
(F_2+i F_3\gamma_5) \frac{i\sigma^{\mu\nu}q_\nu}{2m_N}
\right]
u_{p,\sigma}. \notag
\end{align}
In the original lattice calculation~\cite{Shintani:2005xg}, 
while the CP violation effect on the kinematic coefficient in Eq.~\eqref{eq:2ptcpviol} 
have been correctly taken into account, 
the inadequate definition of these form factors of $\tilde{F}_{1,2,3}$ 
has been used prior to \cite{Abramczyk:2017oxr}.
As a result, all previous lattice results has a 
spurious contributions to the EDFF computed with $\tilde{F}_3$ from the Pauli form factor $F_2$ as 
\begin{align}
\label{eq:correction}
\tilde{F_3} = F_3 -2\alpha F_2.
\end{align}
Thus if the "Old" formula ($\tilde{F_3}$) is used for extracting the nEDM $\tilde{d}_n=\tilde{F}_3(0)/(2m_N)$, 
there is a correction from the spurious mixing, 
which becomes significant when $\alpha$ becomes large.

The inconsistency between "Old" 
and "New" 
formula can be directly and numerically confirmed 
by comparing with computing nEDM from the energy shift method. 
The uniform electric field that preserves translation invariance and the (anti-) periodic boundary 
conditions on a lattice was first introduced in \cite{Izubuchi:2008mu} to study the nEDM, 
and also applied to CP-even quantities such as electric polarizability and magnetic moments 
of the nucleon~\cite{Detmold:2009dx, Detmold:2010ts}.
The uniform background electric field is analytically continued to an imaginary value, 
so that the nucleon energy shift due to nEDM becomes imaginary.
Expanding the two-point function up to the first order in $\theta$ 
we can directly extract the nEDM contribution that is linear in $t$.
For simplicity we only consider the neutral particles, since the correlation function 
of charged particles is more complicated.
To introduce the background electric field we consider the following Euclidean $U(1)$ vector potential 
\begin{align}
A_{x,j}  = 
n_{ij} \Phi_{ij} x_i, 
\quad \quad 
A_{x,i}|_{x_i=L_i-1}  = 
-n_{ij} \Phi_{ij} L_ix_j, 
\end{align}  
where $\Phi_{\mu\nu}=\frac{6\pi}{L_\mu L_\nu}$ is the quantum field flux 
on a plaquette ($\mu\nu$) and $n_{\mu\nu}$ is 
the corresponding number of quanta. 
The (anti-) periodicity on a lattice in both space and time requires the Dirac quantization conditions 
\begin{align}
Q_q \Phi_{\mu\nu} L_\mu L_\nu = 2\pi n_{\mu\nu}, \quad \quad \left(Q_u=\frac23, \ \ Q_d=-\frac13\right)
\end{align}  
The electric field vector is then given as $\vec{E}=(n_{14} \Phi_{14}, n_{24} \Phi_{24}, n_{34} \Phi_{34})$.
The corresponding effective Dirac equation for nucleon field $\tilde{N}$ is given as 
\begin{align}
\left( \Slash{p}-(\tilde{\kappa}+i\tilde{\zeta}\gamma_5)\frac{F_{\mu\nu}}{2}\frac{\sigma^{\mu\nu}}{2m_N}-m_Ne^{-2i\alpha\gamma_5} \right)\tilde{N}=0,
\end{align}  
where $\tilde{\kappa}=\tilde{F}_2(0)$ and $\tilde{\zeta}=\tilde{F}_3(0)$ 
are the effective anomalous magnetic and electric moments 
in the basis of $\tilde{N}$ with $\gamma_5$ mass.
It is obvious that the $\CPviol$ phase $e^{i\alpha \gamma_5}$ 
can be completely rotated away by a field redefinition $N=e^{i\alpha \gamma_5}\tilde{N}$, 
where the two couplings also transform $e^{2i\alpha} (\tilde{\kappa}+i\tilde{\zeta})=(\kappa+i\zeta)$. 
Considering an electric field in $z$-direction $\vec{E}=(0,0,E_z)$ in the rest frame $p^\mu=(\vec{0}, E_s)$,  
we obtain an on-shell solution spinor $u_{E_z,s}$ for the spin polarized along $z$-direction with $s=\pm$, 
which has a spin-dependent energy eigenvalue $E_s=m_N-\frac{\zeta}{2m_N} (s E_z)+O(E_z^2)$. 
From the result 
we see that the resulting energy shift is consistent with the EDFF obtained in the basis of $N$ without $\gamma_5$ mass, 
and $\kappa=F_2(0)$ and $\zeta=F_3(0)$ are the nucleon magnetic and electric dipole moment coefficients. 
By taking the analytic continuation of the electric field $E_z$ to the imaginary value
we obtain the energy shift of a nucleon on lattice as $\tilde{E}_\pm=m_N \pm \delta E$, with $\delta E=-\zeta/(2m_N)iE_z$.
To extract the nucleon energy shift
we calculate the following nucleon two point functions in the presence of the background electric fields 
\begin{align}
C_{{\rm 2pt},\vec{E}}^Q(\vec{p},t) = \sum_{\vec y} e^{-i\vec{p}\cdot \vec{y}} \langle N(\vec{y},t)| \bar{N}(\vec{0},0) 
\sum_x[q(x)] \rangle_{\vec{E}}.
\end{align}  
Using the solution spinor $u_{E_z,s}$ and ignoring excited states 
we can expand $C_{{\rm 2pt},\vec{E}}^Q(\vec{p},t)$ as 
\begin{align}
C_{{\rm 2pt},\vec{E}}^Q(\vec{0},t) &= |Z_N|^2 \sum_{s=\pm} \tilde{u}_{E_z,s} \bar{\tilde{u}}_{E_z,s} \frac{e^{-\tilde{E}_st}}{2\tilde{E}_s} 
\notag \\
\label{eq:eshift}
& =
|Z_N|^2 
\left[ 
\frac{1+\gamma_4}{2}(1-\Sigma^z \delta E t) + i\alpha_5 \gamma_5 
+\Sigma^z \frac{\kappa E_z}{2m^2}\gamma_5
\right] e^{-m_Nt} 
+O(\delta E^2, E_z^2),
\end{align}  
where $\Sigma_z=-i\gamma_x\gamma_y$.
Using the standard ratio method we define an ``effective'' energy shift 
$\zeta_n^{\rm eff}(t)=2m_N d_n=\frac{-2m_N}{E_z} (R(t+1)-R(t))$, 
and $R(t)$ is defined as a ratio of two-point functions with two different spin projections 
\begin{align}
\label{eq:r}
R(t)=\frac{{\rm Tr}[(T_{S_{z+}}-T_{S_{z-}})C_{{\rm 2pt},\vec{E}}^\CPviol(\vec{0},t)]}{2{\rm Tr}[T_pC_{{\rm 2pt},\vec{E}}^\CPviol(\vec{0},t)]},
\end{align}
where we use spin polarization projection operators 
$T_{S_{z\pm}}=\frac{1+\gamma_4}{2} (1\pm \Sigma_z)$, and $T_p = (1+\gamma_4)$.
Thus the nEDM in the energy shift method is independent 
from the parity mixing ambiguity,
from which we numerically verify the consistency with the ``New'' formula ($F_3$). 
For more detail on the analyses, see Ref.~\cite{Abramczyk:2017oxr}. 

\begin{figure}[tbph]
\begin{minipage}{0.5\hsize}
\vspace{3mm}
\begin{center}
\includegraphics[clip,width=0.95\textwidth]{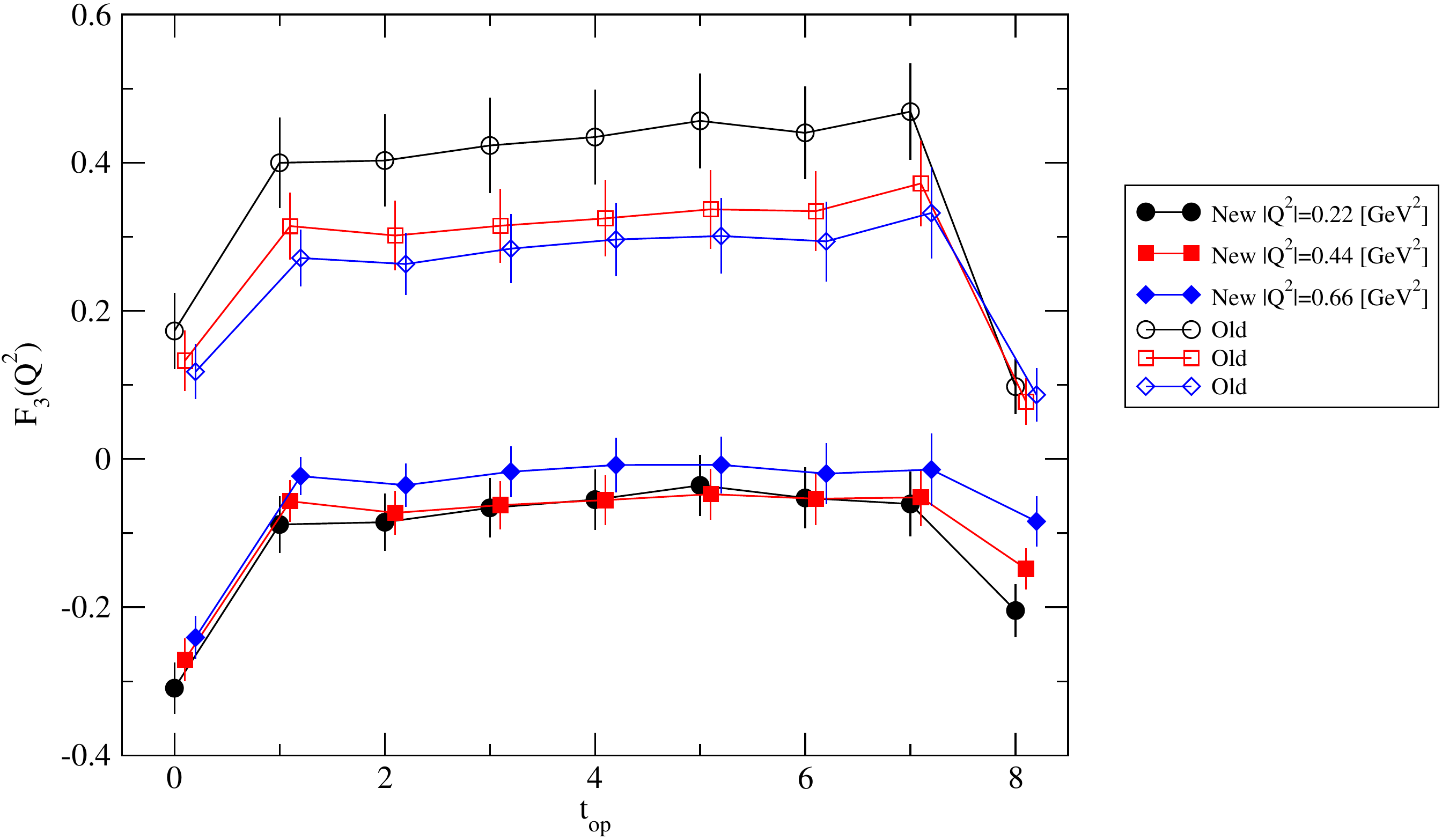} 
\end{center}
\vspace{-3mm}
\caption{\label{fig:theta_t8} 
Plateaus for the nucleon EDFF $F_3(Q^2)$ from QCD $\theta$ term
for $t_{op}=8$. The results with "Old" include spurious mixing with the $F_2$.
Results are shown for a lattice ensemble of domain wall fermion configurations 
of $24^3\times64$ for $m_\pi=330$ MeV. }
\end{minipage}
\ \ \ \ \ \ 
\begin{minipage}{0.5\hsize} 
\begin{center}
\centering
\includegraphics[clip,width=0.7\textwidth]{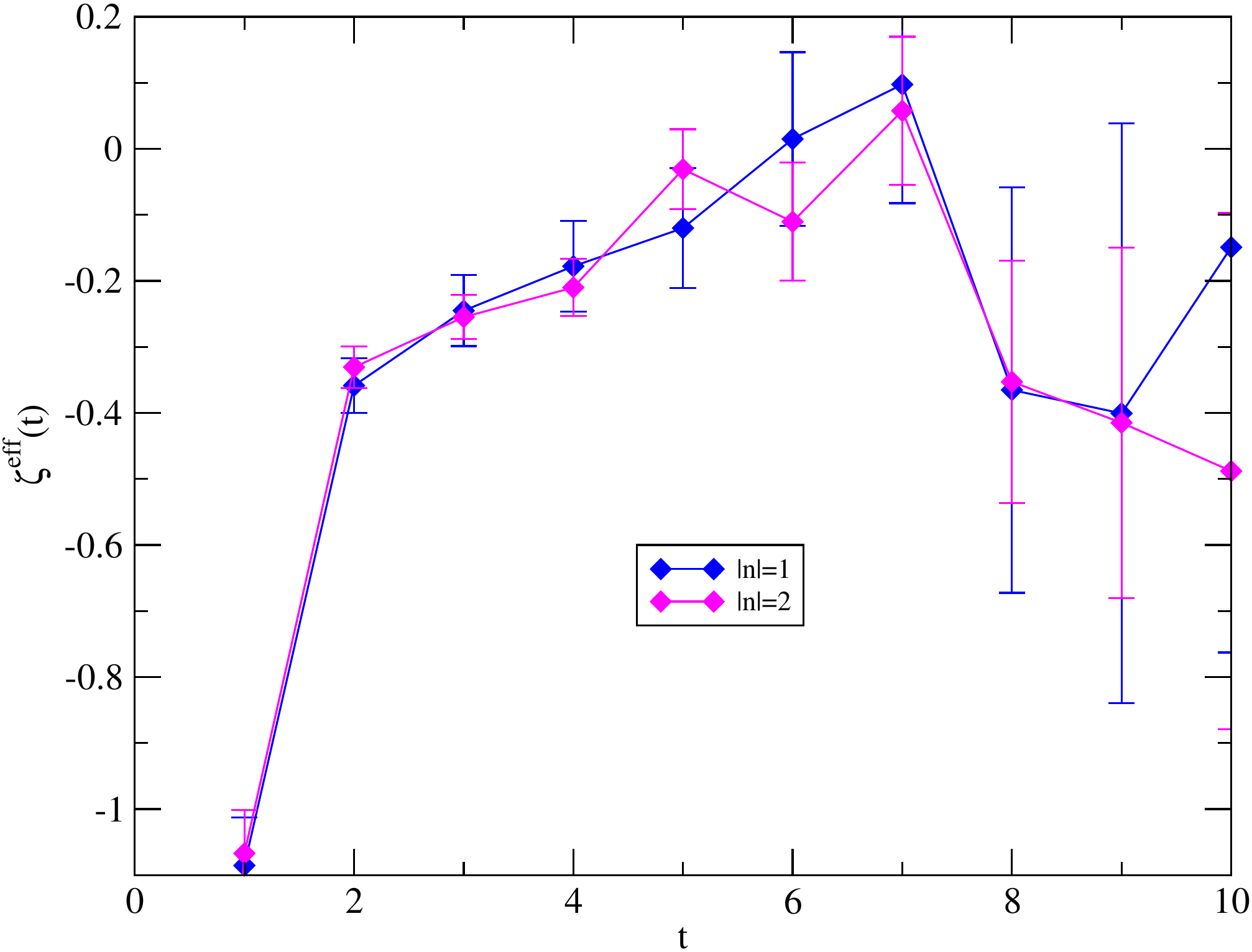} 
\end{center}
\vspace{-5mm}
\caption{\label{fig:theta_BG} 
Effective nEDM $\zeta_n^{\rm eff}$ from QCD $\theta$ term.
Results are shown for a lattice ensemble of domain wall fermion configurations 
of $24^3\times64$ for $m_\pi=330$ MeV. }
\end{minipage}
\end{figure}

We show some nEDM results in comparison between the form factors and the energy shift methods. 
Fig.~\ref{fig:theta_t8} shows how the spurious mixing affects the result for the $\theta$-EDM. 
As shown in the figure, the magnitudes of $F_3(Q^2)$ with "New" are smaller than the "Old" values.
We also show the result for nEDM $\zeta_n^{\rm eff}$ computed from the effective energy shift in Fig.~\ref{fig:theta_BG},
where we computed with two values of flux quanta $n=n_{34}=\pm1$ and $\pm2$.
A plateau for the effective energy $\zeta_n^{\rm eff}$ at $t=4\sim 7$ 
can be obtained in both electric fields of $|n|=1$ and $2$. 
Comparing the form factor method we obtain a consistent result only if we use the ``New'' formula.
This result is also consistent with naively corrected data using the previously reported values prior to~\cite{Abramczyk:2017oxr}.   
Since we obtain a small but non-zero signal $|d_n^\theta| \lesssim 0.1 \frac{e}{2m_N}$ at $m_\pi=330$ MeV,
we estimate an extrapolated value to the physical point based on a naive scaling of the ChPT expectation $d_n \sim m_q \sim m_\pi^2$,
which suggests  
that the signal becomes weaker as the quark mass is approaching to the physical point 
and we would obtain a smaller value of $|d_n^\theta(m_\pi^{\rm phys})| \lesssim 0.02 \frac{e}{2m_N}$.
From our findings in order to promote a direct calculation of nEDM at the physical point, 
various noise reduction techniques that work in particular for gluon operator are
required in addition to a significant increase in statistics.

%%%%%%%%%%%%%%%%%%%%%%%%%%%%%%%%%% 
%%%%%%%%%%%%%%%%%%%%%%%%%%%%%%

\section{Noise reduction technique for $\theta$-EDM}
\begin{figure}[!htbp]
\begin{minipage}{.4\textwidth}
\centering
\includegraphics[clip,width=0.6\textwidth]{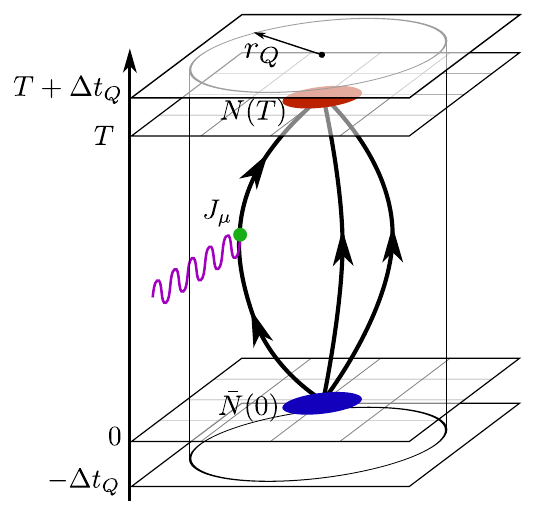} 
\end{minipage}~\hspace{.05\textwidth}~
\begin{minipage}{.5\textwidth}
\caption{\label{fig:Qslab} 
Truncated sampling of the topological charge density \eqref{eq:Qtopo_cuts}
for reducing the noise 
in the CP-odd nucleon correlation functions \eqref{eq:c2pt} and \eqref{eq:c3pt}.
The correlation with the points outside $V_Q$ is expected to be suppressed 
but gives a large noise. 
}
\end{minipage}
\end{figure}

As explained in the previous section, 
$\theta$-induced nEDM would be extremely challenging in the physical point, 
because the topological charge fluctuation dominates the large statistical noise 
growing with lattice volume $V_4$ as $\delta Q^2 = \langle Q^2 \rangle \propto V_4$,
while the signal becomes small. 
Since the global topological charge is zero on average because our QCD action is CP-even, 
$F_3(Q^2)$ is the signal of the correlation between the gluon operator and the fermionic (nucleon) functions
as shown in Eq.~\eqref{eq:c3pt}.
Thus it has been originally suggested that truncation of the topological charge sum at a large distance
from the nucleon position
can reduce fluctuations of the nEDM~\cite{Shintani:2015vsx} 
(also known as a cluster decomposition of disconnected diagrams~\cite{Liu:2017man}), 
since contributions of $Q$ at large distance may be neglected in computing the nEDM, 
while its correlation has a large noise which is not suppressed with space-time distance
due to the global nature of the topological charge. 
To extend this method, 
in Ref.~\cite{Syritsyn:2019vvt} 
we consider a generalized reduced topological charge density which separately restrict time and space to a cylindrical volume $V_Q$,
\begin{align}
\label{eq:Qtopo_cuts}
\tilde{Q}(\Delta t_Q,r_Q) = \frac1{16\pi^2} \sum_{x\in V_Q}{\rm Tr}\big[ \hat{G}_{\mu\nu} \tilde{\hat{G}}_{\mu\nu}\big]_{x}\,,
\quad (\vec x,t)\in V_Q :
\left\{
\begin{array}{l} 
  |\vec x-\vec x_0|\le r_Q\,, \\
  t_0 - \Delta t_Q < t < t_0 + t_{\rm sep} + \Delta t_Q\,,
\end{array}
\right.
\end{align}
where $t_0$ and $t_0+t_{\rm sep}$ are the positions of the nucleon source and sink.
This setup is illustrated in Fig.~\ref{fig:Qslab}, 
where the three-point functions are inside entirely the region $V_Q$ 
and the truncation in $t$-direction for $\tilde{Q}$ is symmetric with respect
to the nucleon sources and sinks for both two- and three-point functions.
We also set $\vec{x}_0$ to the nucleon source position 
to further reduce the noise at large distances. 
We should note that a spatial restriction may introduce another bias for nEDM. 
In order to avoid such ambiguity, 
the convergence with $r_Q$ and $\Delta t_Q$ in Eq.~\eqref{eq:Qtopo_cuts} 
must be verified at each nucleon momenta, 
especially in computing the $Q^2$ dependence of the EDFFs.

%%%%%%%%%%%%%%%%%%%%%%%%%%%%%%%%%% 
%%%%%%%%%%%%%%%%%%%%%%%%%%%%%%

\begin{figure}[!ht]
\centering
\includegraphics[clip,width=15cm]{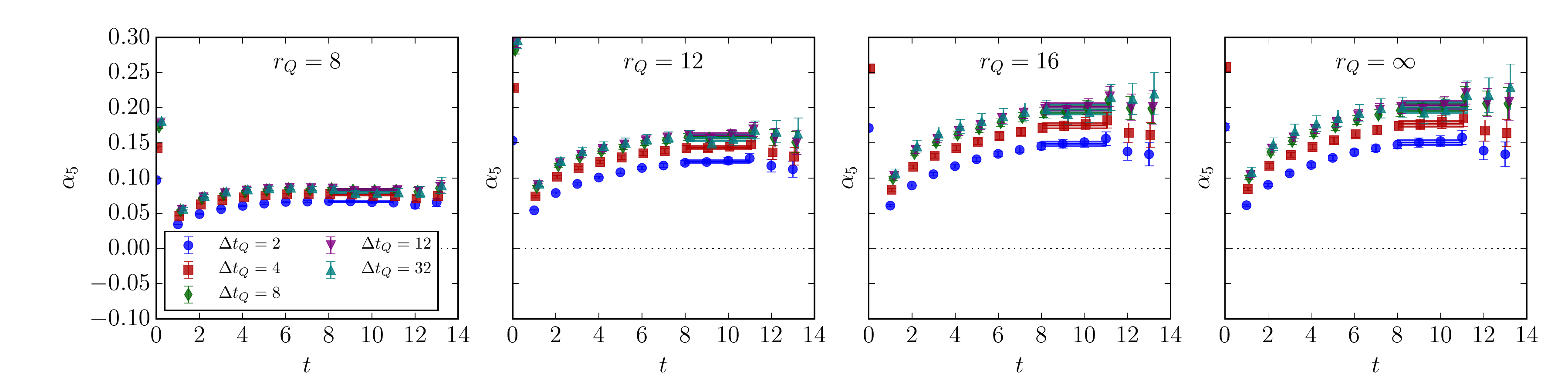} 
\caption{\label{fig:alpha5} 
Dependence of the spatial and temporal cuts ($r_Q$, $\Delta t_Q$) 
in the reduced topological charge \eqref{eq:Qtopo_cuts}
on the nucleon parity mixing angle $\alpha_5$.  
}
\end{figure}

\begin{figure}[!ht]
\centering
\includegraphics[clip,width=16cm]{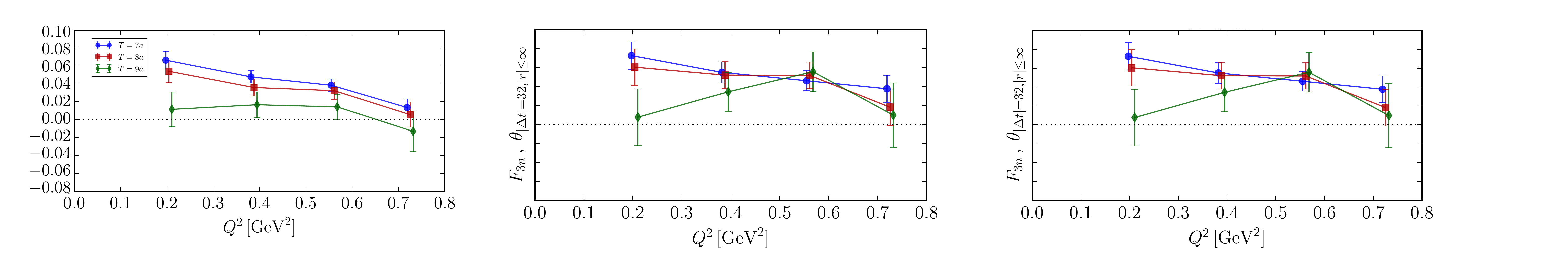} 
\caption{\label{fig:edff330} 
Neutron EDFF induced by $\theta$-term and its dependence of the spatial and temporal cuts ($r_Q$, $\Delta t_Q$) 
at $m_\pi=330$ MeV.   
}
\end{figure}

\begin{figure}[!ht]
\centering
\includegraphics[clip,width=14cm]{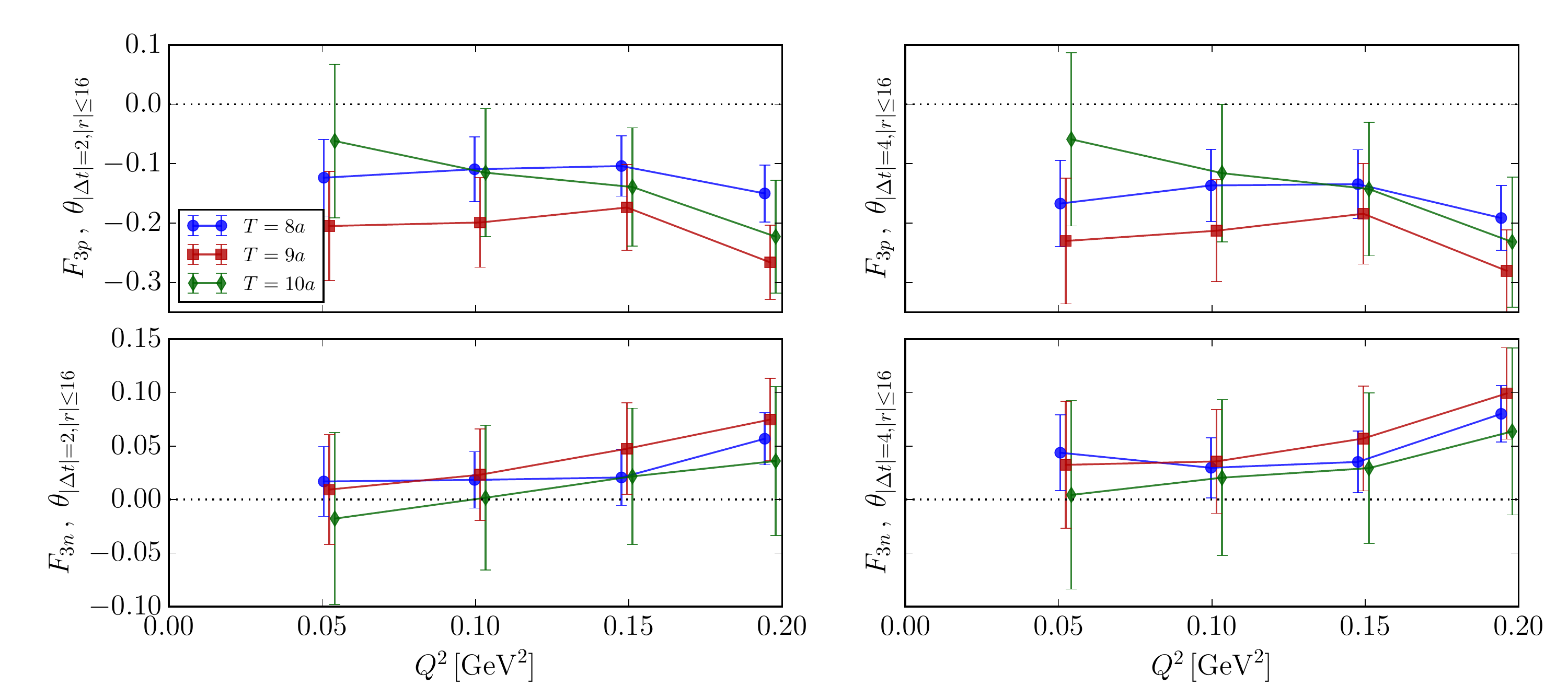} 
\caption{\label{fig:edffphys} 
Preliminary results for the proton and neutron EDFF induced by $\theta$-term 
and its dependence of the spatial and temporal cuts ($r_Q$, $\Delta t_Q$) 
at $m_\pi=139$ MeV \cite{Syritsyn:2019vvt}.   
}
\end{figure}

We use the lattice QCD ensemble of dynamical domain wall fermion with heavy pion mass $m_\pi = 330$ MeV. 
We calculate 64 low-precision and 1 high-precision samples using the AMA sampling method \cite{Blum:2012uh}.
The topological charge density is calculated 
using the ``5-loop-improved'' field strength $\hat{G}_{\mu\nu}$ \cite{deForcrand:1997esx}
with the gradient flow ($t_{gf}=8a^2$).
The $r_Q$ and $\Delta t_Q$ dependence of the mixing angle $\alpha$ and the EDFF for the neutron (connected diagram only) 
are shown in Figs.~\ref{fig:alpha5} and \ref{fig:edff330}, 
where we observe error reduction for smaller values of $r_Q$ and $\Delta_Q$
and convergence for $r_Q \geq 16$, $\Delta t_Q \geq 8$ in both $\alpha$ and $F_3(Q^2)$. 
We have also performed a calculation using ensembles at the physical point 
on $48^3 \times 96$ lattice with 33,000 statistics.
Unfortunately we have found no signal for neutron EDFFs (See Fig.~\ref{fig:edffphys}), 
and the results are consistent with zero with the statistical uncertainty. 
Our result of signal-to-noise ratio $\sim 0.2$ at $m_\pi=330$ MeV indicates that 
the expected signal-to-noise ratio at the physical point 
has to be improved by a factor of $\sim$10 which requires $\times O(100)$ 
more statistics to confirm the existence of the strong CP problem. 

%%%%%%%%%%%%%%%%%%%%%%%%%%%%%%%%%% 
%%%%%%%%%%%%%%%%%%%%%%%%%%%%%%

%%%%%%%%%%%%%%%%%%%%%%%%%%%%%%%%%% 
%%%%%%%%%%%%%%%%%%%%%%%%%%%%%%

%%%%%%%%%%%%%%%%%%%%%%%%%%%%%%%%%% 
%%%%%%%%%%%%%%%%%%%%%%%%%%%%%%

\begin{figure}[!ht]
\centering
\includegraphics[clip,width=6cm]{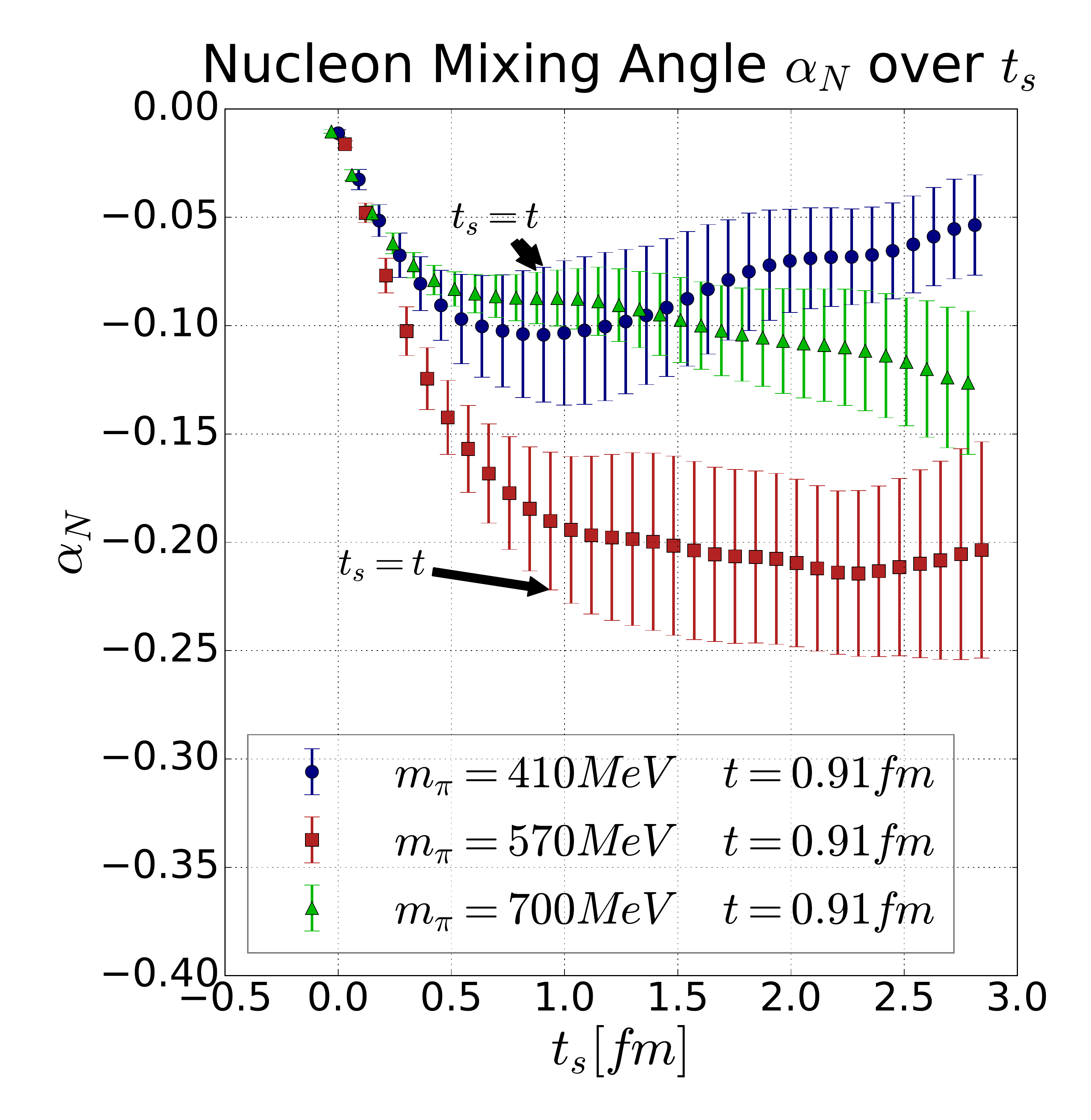} \quad \quad
\includegraphics[clip,width=6cm]{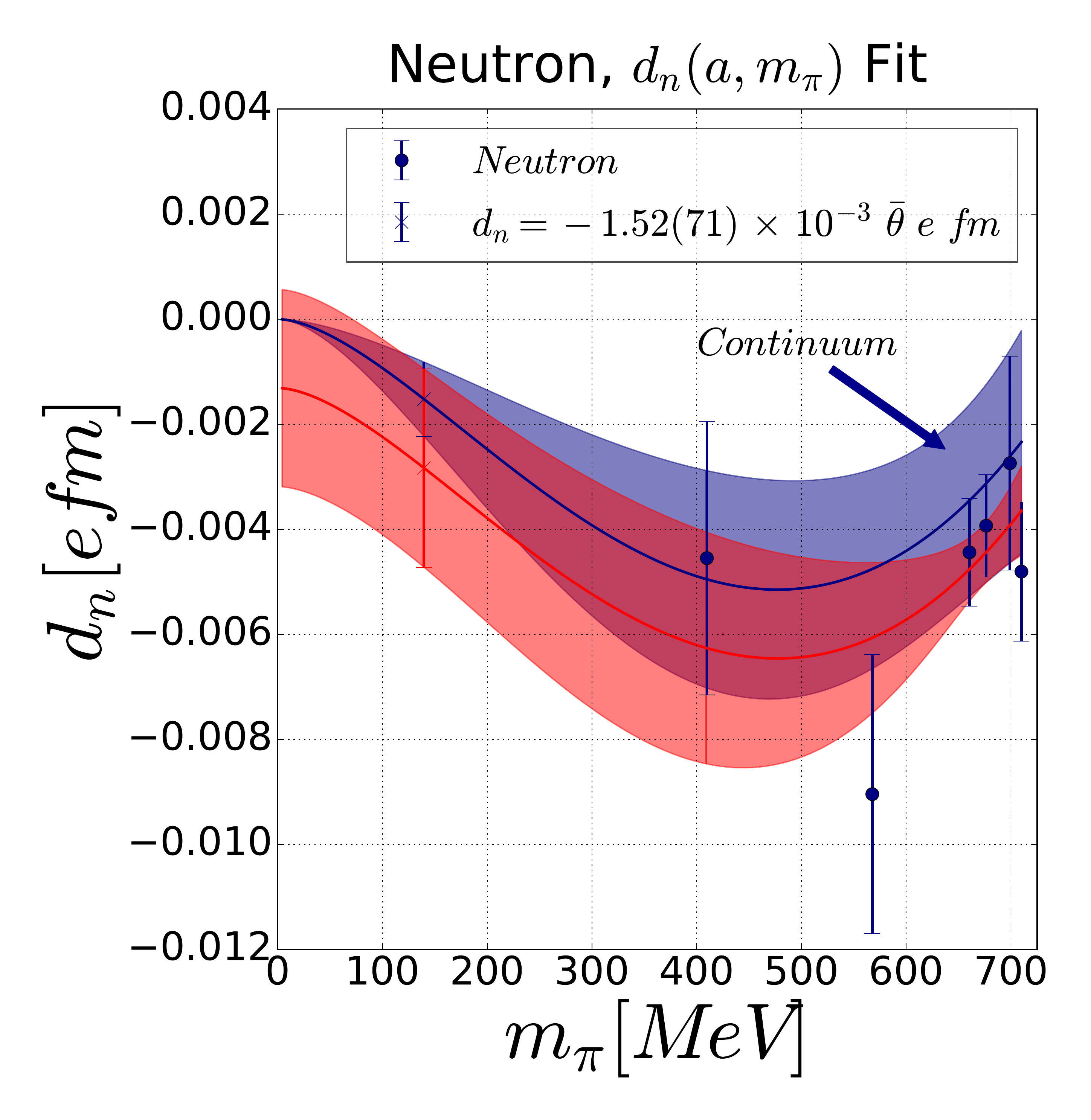} 
\caption{\label{fig:Dragos} 
Results presented in Ref.~\cite{Dragos:2019oxn}.
(Left) The improved nucleon mixing angle $\alpha$ plotted against the sum parameter $t_s$. \ 
(Right) A chiral and continuum extraplation of nEDM $d_n^\theta$ plotted 
as a function of $m_\pi$. The bands are the fit results. }
\end{figure}

A systematic analysis of the truncation method based on spectrum decomposition is proposed in Refs.~\cite{Dragos:2018uzd, Dragos:2019oxn}.
This study uses a truncation of the topological charge density operator summed over spatial directions $\bar{Q}$ defined as
\begin{align}
\label{eq:qop}
\bar{Q}(t_Q)=\int d^3x q(x,t_Q), \quad \quad 
q(x,t_Q)=\frac{1}{16\pi^2} G\tilde{G}(x,t_Q)
\end{align}
where the topological charge density $q(x,t_Q)$ is again calculated using the gradient flow method.
To calculate the nucleon mixing angle,  the following modified two-point function 
\begin{align}
\Delta C_{\rm 2pt}(t,t_Q)=\langle {\rm T}\{N(T)\bar{Q}(t_Q)\bar{N}(0)\} \rangle,
\end{align}
is considered.
The spectrum decomposition for $\Delta C_{\rm 2pt}(t,t_Q)$ 
reveals its $t_Q$ dependence and provide a systematic way to estimate the truncation error. 
For example, in the case of $0<t_Q<t$, its spectral decomposition has the form
\begin{align}
\label{eq:sd1}
\Delta C_{\rm 2pt}(t,t_Q) = 
\langle N(t)\bar{Q}(t_Q)\bar{N}(0)\rangle &\sim 
\sum_{n,m} e^{-E_n (t-t_Q)-E_m t_Q} \langle 0|N|n\rangle \langle n|\bar{Q}|m\rangle \langle m|\bar{N}|0\rangle
\notag \\
& \sim \sum_{m\ne n} \cosh{\left(\Delta m_{mn}(t_Q-t/2)\right)}, 
\end{align}
where $\Delta_{mn}=E_m-E_n$.  
Note that two states of $n$ and $m$ should have different intrinsic parities, 
otherwise the contribution vanishes  
since the matrix element should be zero, 
e.g., $\langle n |\bar{Q}|n\rangle=0$ 
in CP-even QCD vacuum
if $n$ has even or odd parity.
On the other hand, in the case of $t<t_Q$, it has 
\begin{align}
\label{eq:sd2}
\Delta C_{\rm 2pt}(t,t_Q) = 
\langle \bar{Q}(t_Q)N(t)\bar{N}(0)\rangle &\sim 
\sum_{n,m} e^{-E_n t_Q-E_m t} \langle 0|\bar{Q}|n\rangle \langle n|N|m\rangle \langle m|\bar{N}|0\rangle
\notag \\
& \sim \sum_{n'} e^{-E_{n'}t_Q}, 
\end{align}
where the state $n'$ should be a P-odd state 
that couples to a nucleon state with a non-zero value of the matrix element $\langle 0|\bar{Q}|n'\rangle \ne 0$.  
From the spectrum decomposition 
we see that the nucleon mixing angle $\alpha$ is a mixing parameter between the ground state nucleon and CP-odd excited states.
To see the truncation artefacts,  
the authors of Refs.~\cite{Dragos:2018uzd, Dragos:2019oxn}
consider the partial summed two-point correlation function 
$\displaystyle C^{\bar Q}_{\rm 2pt}(t_s) = \sum_{t_Q=-t_s}^{t_s} \Delta C_{\rm 2pt}(t,t_Q)$.
From Eqs.~\eqref{eq:sd1} and \eqref{eq:sd2} 
its asymptotic form behaves like $\displaystyle  C^{\bar Q}_{\rm 2pt}(t_s)=A+Be^{-E t_s}$ for $t_s \gtrsim t$,
where the contribution for large $t_s$ is expected to be exponentially suppressed. 
This is numerically checked as shown in the left panel of Fig.~\ref{fig:Dragos},
where a plateau is obtained for $t_s \sim t$ and the contributions from $t_s >t$ 
seem to be below the statistical fluctuation. 
Neglecting unnecessary noise from $t_s \gtrsim t$,  
an improvement of $\alpha$ up to a factor 2 
is obtained.
The same spectrum decomposition can be applied to the modified nucleon three point function 
$\Delta C^{\bar Q}_{\rm 3pt}(t,t_Q,t_{\rm op})=\langle {\rm T}\{N(T)J^\mu(t_{\rm op})\bar{Q}(t_Q)\bar{N}(0)\} \rangle$, 
where a fit analysis using its asymptotic form 
with the $\alpha$-improvement  
yields a factor of $2 \sim 3$ times increases in the signal-to-noise ratio for EDFF $F_3(Q^2)$.
The right panel of Fig.~\ref{fig:Dragos} shows a chiral (and continuum) extrapolation for $d_n^\theta$ 
using six ensembles with $m_\pi >410$ MeV for several lattice spacings.
A ChPT fit ansatz $d_n^\theta \sim m_q \sim m_\pi^2$ is used 
to further constrain the nEDM towards the chiral limit, 
which yields $d_n^\theta=-0.00152(71) \theta$ $e$ fm at the physical point with $\sim$ 2 $\sigma$ deviation from zero. 
This result is consistent with the aforementioned naively scaled value $|d_n^\theta(m_\pi^{\rm phys})| \lesssim 0.02 \frac{e}{2m_N}$ \cite{Syritsyn:2019vvt}. 
Their fitted data, however, are in heavy pion mass region and do not clearly show the chiral behavior $d_n^\theta\propto m_\pi^2$,  
so that the fit results seem to be less convincing. 
To avoid model dependence, more accurate results near the physical point are needed.

%%%%%%%%%%%%%%%%%%%%%%%%%%%%%%%%%% 
%%%%%%%%%%%%%%%%%%%%%%%%%%%%%%

\begin{figure}[!ht]
\centering
\includegraphics[clip,width=6.2cm]{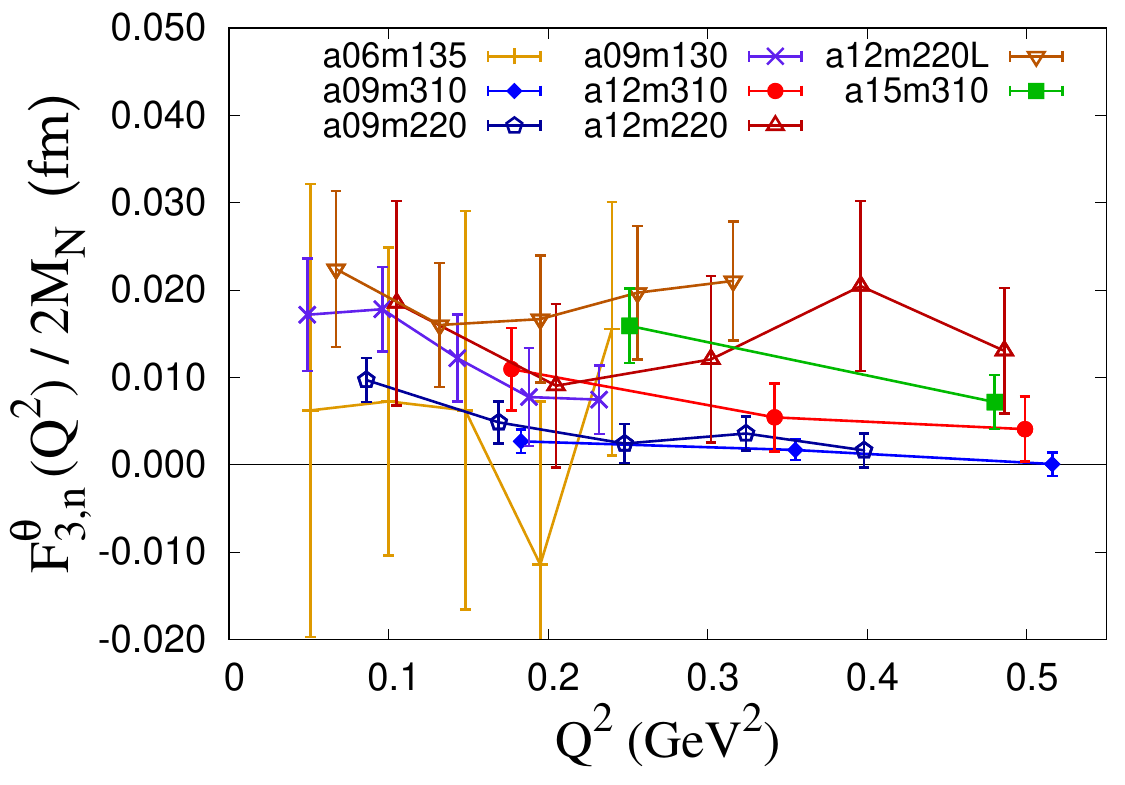} \quad 
\includegraphics[clip,width=6.5cm]{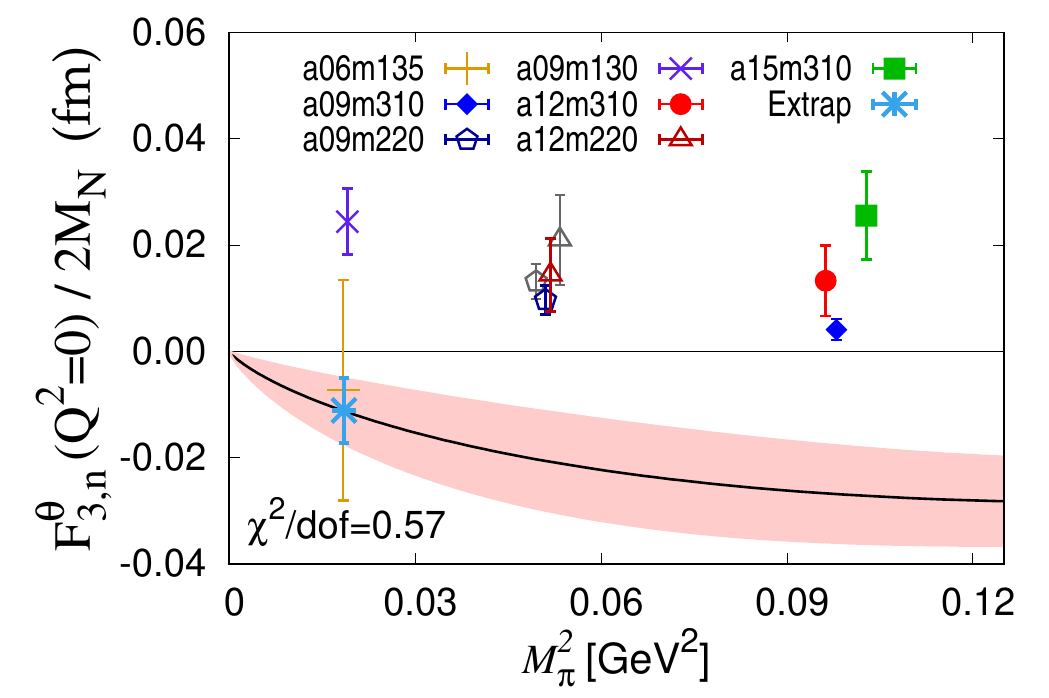} 
\caption{\label{fig:Yoon} 
Preliminary results presented in Ref.~\cite{Yoon:2020soi}.
(Left) The $Q^2$ dependence of $F_3(Q^2)/(2m_N)$. \ 
(Right) A chiral and continuum extraplation of nEDM $d_n^\theta$ plotted 
as a function of $m_\pi^2$. }
\end{figure}

New results near the physical point ($m_\pi=130$-$310$ MeV) are presented in a parallel talk by Boram Yoon~\cite{Yoon:2020soi}. 
The topological charge is calculated using the $O(a^4)$-improved field strength \cite{BilsonThompson:2002jk}
with the gradient flow on MILC HISQ ensembles with $a=0.06-0.15$ fm. 
This study also uses the truncation method in $t$-direction.
While the convergence properties in partial sum 
can be seen in both two- and three- point functions, 
due to the slow convergence at the physical point 
no significant improvement is observed. 
It is, however, remarkable that the number of measurements is $O$(100k), 
which gives a statistically significant signal for $F_3(Q^2)$ with non-zero $Q^2$ 
near the physical point (see the left panel of Fig.~\ref{fig:Yoon}).
It is also reported that a variance reduction technique introduced in \cite{Bhattacharya:2018qat} 
has about $25$\% error reduction.
In this calculation the excited state contamination is removed by using the two-state fits with multiple source-sink separations for 
each momentum $Q^2$. 
The results for $F_3(Q^2)$ at heavier $m_\pi$ with non-zero $Q^2$ 
are consistent with the previous lattice results \cite{Syritsyn:2019vvt, Dragos:2019oxn}.
The result from the chiral (interpolation) and continuum extrapolation fit for $d_n^\theta$ 
are also presented in the right panel of Fig.~\ref{fig:Yoon}, 
which yields a non-zero signal of $|d_n^\theta|=0.011(6) \theta$ $e$ fm at the physical point.
While this result is also consistent with estimates from ChPT analysis and the QCD sum rules~\cite{Engel:2013lsa},
it is not sufficient to constrain the $\theta$ parameter. 
We also notice 
that even though the values of $d_n^\theta$ at finite $a$ are all positive except for the data at $m_\pi=135$ MeV, 
these values in the continuum limit become negative,
which may indicate 
a sizable discretization effect on $d_n^\theta$.
In addition, there is an increasing tendency of $F_3(Q^2)$ towards $Q^2, m_\pi^2 \to 0$ but with larger error.
A leading order ChPT fit may not be suitable for $d_n^\theta$ in the simulation mass region,  
and understanding the $Q^2$ dependence of $F_3(Q^2)$ would be 
important to precisely determine $d_n^\theta$ at the physical point. 

We have shown in this section that there are several ongoing studies 
of the $\theta$-induced nEDM. Even having $O$(100k) statistics with 
employing noise reduction techniques at the physical point, 
it still is not sufficient to constrain the $\theta$ parameter due to the large 
fluctuation of the topological charge density, which should become even worse 
when approaching to the continuum and larger volume limit. 
There also are several systematic uncertainties due to finite lattice spacing and 
$Q^2\to 0$ extrapolation in EDFF $F_3(Q^2)$, 
which need to be further explored.  
Thus $\theta$-induced EDM at the physical point will be extremely challenging 
and will require more special techniques 
that work well in particular for gluon operators.
In the next section, we would like to propose a new approach 
using a matrix element of the nucleon with background electric field. 
Since this approach is based on the energy shift method, 
we can directly obtain $d_n^\theta$ without $Q^2$ extrapolation,
which may be potentially advantageous over the form factor method.

%%%%%%%%%%%%%%%%%%%%%%%%%%%%%%%%%% 
%%%%%%%%%%%%%%%%%%%%%%%%%%%%%%

\section{New approach based on the matrix element with the background electric field}
The idea is to simply apply the truncation technique to 
the energy shift method in the presence of the background electric field, 
in which we find that the energy shift $\delta E$ in Eq.~\eqref{eq:eshift} 
is given as a nucleon matrix element of the topological charge density operator $\bar{Q}$ 
in Eq.~\eqref{eq:qop}. 
Throughout the section, the state's momenta are set to zero and 
arguments of $\vec{p}$ are suppressed.
Performing the spectrum decomposition of $\Delta C_{\rm 2pt}(t,t_Q)$ as in Eq.~\eqref{eq:sd1} but now in the presence of 
the background electric field, 
we obtain the following result for $t>t_Q$ as 
\begin{align}
\label{eq:sdbgef1}
\Delta C_{{\rm 2pt},\vec{E}}(t,t_Q) &= 
\langle N(t)\bar{Q}(t_Q)\bar{N}(0)\rangle_{\vec{E}} \notag \\
&= \sum_{n,m} e^{-E_n (t-t_Q)-E_m t_Q} 
\langle 0|N|n,\vec{E}\rangle \langle n,\vec{E}|\bar{Q}|m,\vec{E}\rangle \langle m,\vec{E}|\bar{N}|0\rangle
\notag \\
&\sim |Z_{N_+}|^2 e^{-m_{N_+} t} \langle N_+,\vec{E}|\bar{Q}|N_+,\vec{E}\rangle, 
\end{align}
where $|N_+,\vec{E}\rangle$ is the ground state nucleon in the presence of the background electric field.
We should note that in contrast to Eq.~\eqref{eq:sd1}, there is a leading order contribution 
from the ground state nucleon given as a matrix element $\langle N_+,\vec{E}|\bar{Q}|N_+,\vec{E}\rangle$.  
Again taking the partial summation of $\Delta C_{{\rm 2pt},\vec{E}}(t,t_Q)$ over sink-source separation $t$,
we obtain 
\begin{align}
\label{eq:sdbgef2}
C_{{\rm 2pt},\vec{E}}^{\bar Q}(t)=\sum_{t_Q=0}^t \Delta C_{{\rm 2pt}, \vec{E}}(t,t_Q) \sim 
|Z_{N_+}|^2 e^{-m_{N_+}t}(t \langle N_+,\vec{E}|\bar{Q}|N_+,\vec{E}\rangle), 
\end{align}
where the matrix element is given the coefficient of linear in $t$. 
Comparing Eq.~\eqref{eq:eshift} with Eq.~\eqref{eq:sdbgef2}, 
we find that the matrix element should correspond to the energy shift 
\begin{align}
\label{eq:me1}
\langle N_+,\vec{E}|\bar{Q}|N_+,\vec{E}\rangle = - \frac{\zeta}{2m_N} \bar{u}\left[ \vec{\Sigma}\cdot \vec{E} \right]u + O(\vec{E}^2).
\end{align}
This formula is analogous to the leading order energy correction in the perturbation theory of quantum mechanics, 
c.f., $\Delta E_n= \langle n| \Delta \hat{H} |n \rangle$ for Hamiltonian $\hat{H}=\hat{H}_0+\Delta \hat{H}$.
In this case we regard the electric field as a perturbation 
in addition to the CP-odd Hamiltonian ($\theta$-term), 
and also take into account a leading order perturbation effect on the state $|N_+,\vec{E}\rangle$. 
Even without $\theta$-term, 
the ground state $|N_+,\vec{E}\rangle$ 
could mix with CP-odd states \cite{Baym:2016lyf} due to the background electric field as 
\begin{align}
\label{eq:sm}
|N_+,\vec{E}\rangle &= |N_+\rangle +c \vec{E}\cdot\vec{D} |N_-\rangle + \cdots,
\end{align}
where $|N_+\rangle$ is the parity-even ground state nucleon in (lattice) CP-even vacuum.
The leading order correction should come from a parity-odd nucleon $|N_-\rangle$ 
with an overlap coefficients of $c$ and the electric field $\vec{E}$ and 
an expectation value of the dipole operator $\vec{D}$ ~\cite{Baym:2016lyf}.
Substituting Eq.~\eqref{eq:sm} into Eq.~\eqref{eq:me1}, we obtain 
\begin{align}
\label{eq:me2}
\langle N_+,\vec{E}|\bar{Q}|N_+,\vec{E}\rangle &= 
\vec{E}\cdot\vec{D} \langle N_+|\bar{Q}|N_-\rangle + ({\rm c.c.})+\cdots,
\end{align}
where we note $\langle N_\pm|\bar{Q}|N_\pm\rangle=0$ due to parity symmetry,  
since we consider the electric field as a perturbation so that $|N_\pm\rangle$ is defined 
in CP-even QCD vacuum.   
Thus the contribution of the matrix element $\langle N_+|\bar{Q}|N_-\rangle$ 
is exactly the same as the parity mixing effect 
that appears in calculation of $\alpha$ in Eq.~\eqref{eq:sd1}.
In the perturbation theory point of view, 
the EDM is an interplay of the electric field and the CP-odd operator 
both in the first order perturbation. 
Using the standard ratio method as in Eq.~\eqref{eq:r} 
and the relation to the matrix element in Eq.~\eqref{eq:me1}, 
we obtain the modified energy shift formula 
\begin{align}
\label{eq:r2}
R_{S_{z\pm}}(t,t_Q)=\frac{{\rm Tr}[T_{S_{z\pm}} \Delta C_{{\rm 2pt},\vec{E}}(t,t_Q)]}{{\rm Tr}[T_p C_{2pt,\vec{E}}(t)]} \to 
\mp \frac{\zeta}{2m_N}E_z,  \quad \quad (t \to \infty), 
\end{align}
From this formula, it is clear that we do not need to extend $t_Q$ outside the sink-source position, 
since there is no other term that is proportional to $t$ in $t_Q>t$. 
In fact the contributions from $t_Q >t$ should be excited state contaminations that 
should disappear in the limit $t \to \infty$. 
Thus, without $Q^2$ extrapolation, the EDFF $F_3(0)$ can be directly extracted from the ratio by dividing by the electric field
as $\displaystyle |F_3(0)|=\lim_{t \to \infty}\frac{2m_N |R_{S_{z\pm}}(t,t_Q)|}{|E_z|}$. 

\begin{figure}[htb]
\centering
\includegraphics[clip,width=0.47\textwidth]{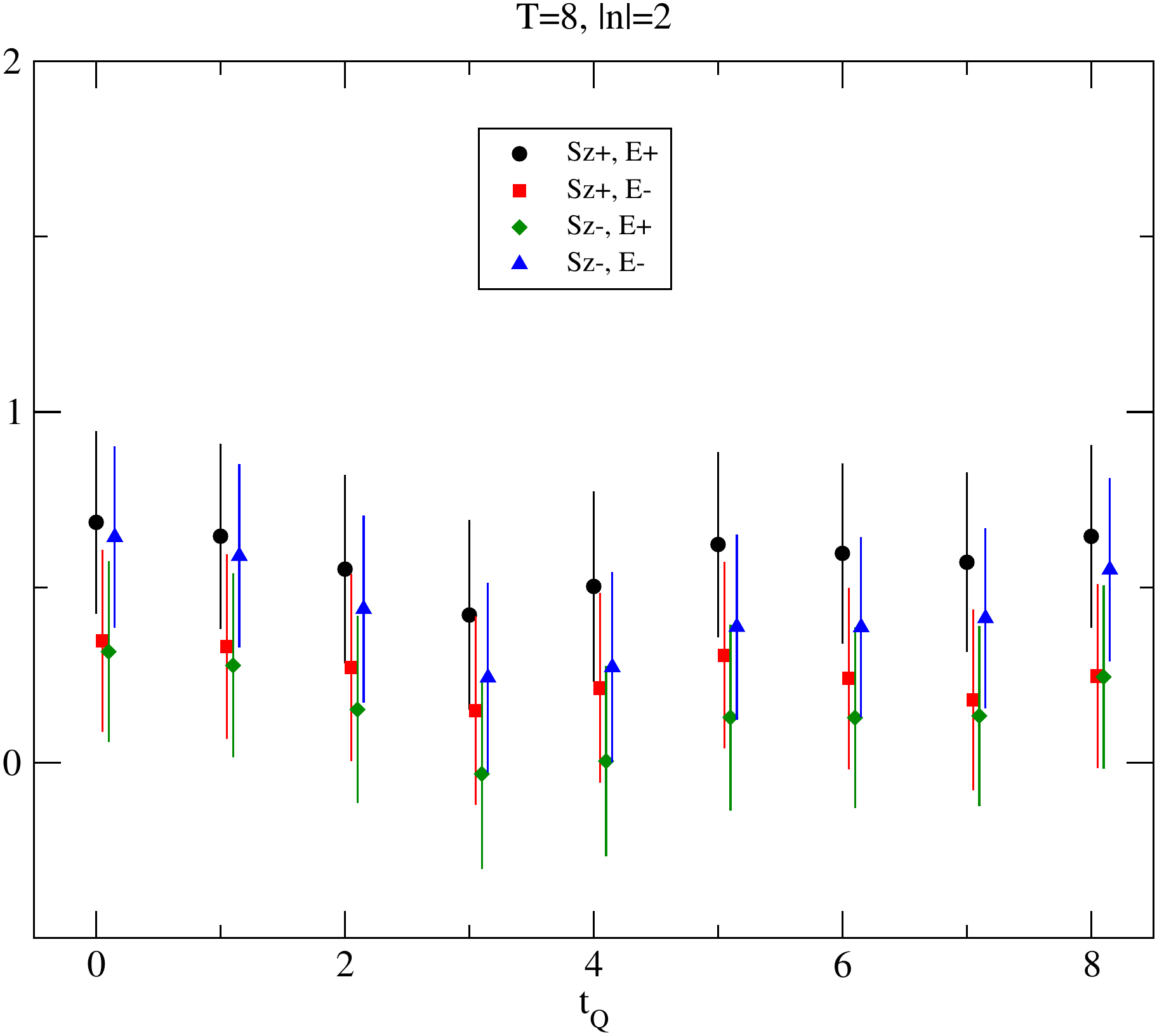} \quad \ 
\includegraphics[clip,width=0.49\textwidth]{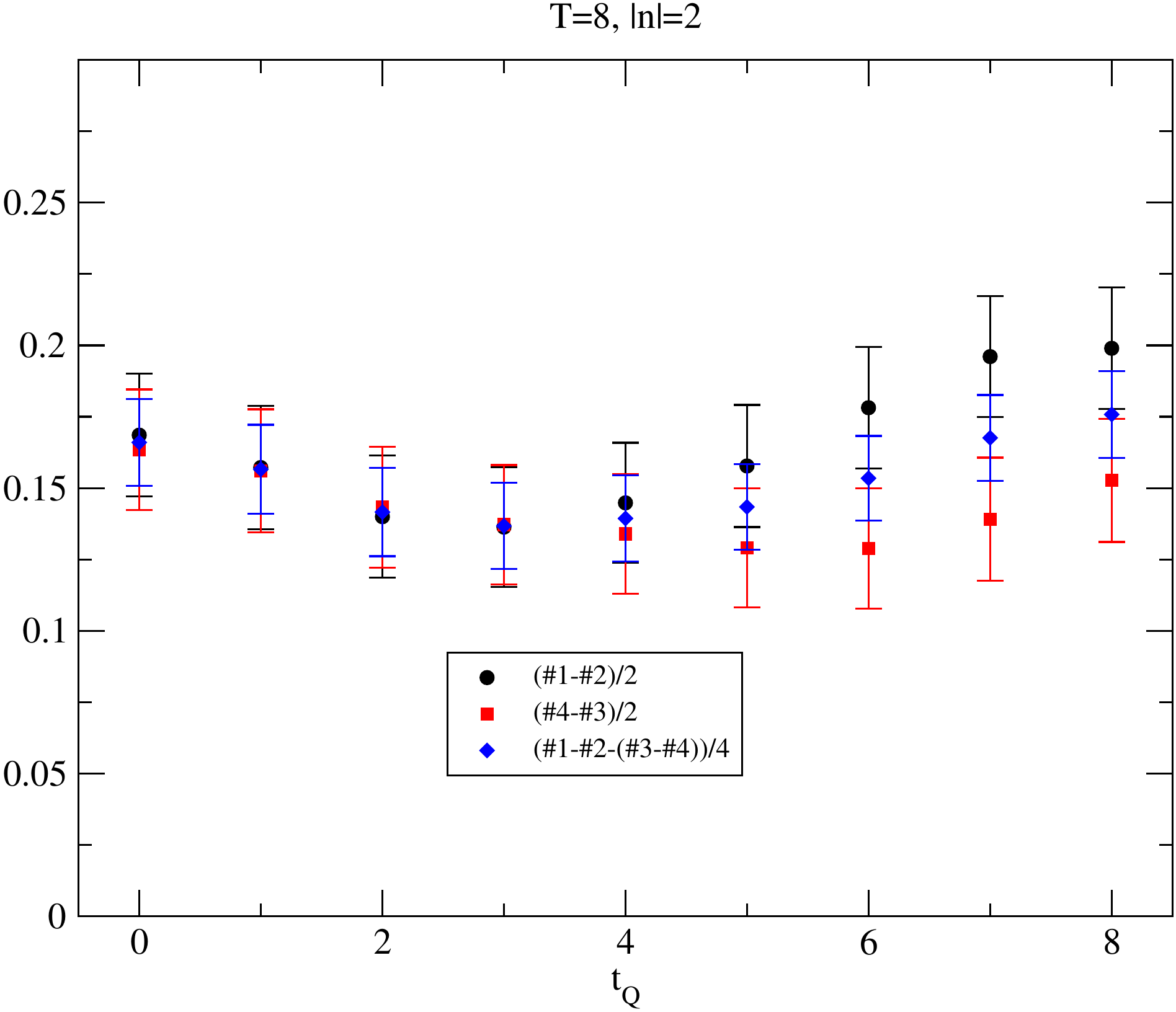}  
\caption{
Preliminary results of $F_3(0)$ from the matrix element approach. 
Results are obtained on a $24^3\times 64$ lattice with $m_\pi=330$ MeV.
4 different components (Left) and their linear combinations 
for obtaining better signals (Right).
\label{fig:component}}
\end{figure}

\begin{figure}[htb]
\centering
\includegraphics[clip,width=0.49\textwidth]{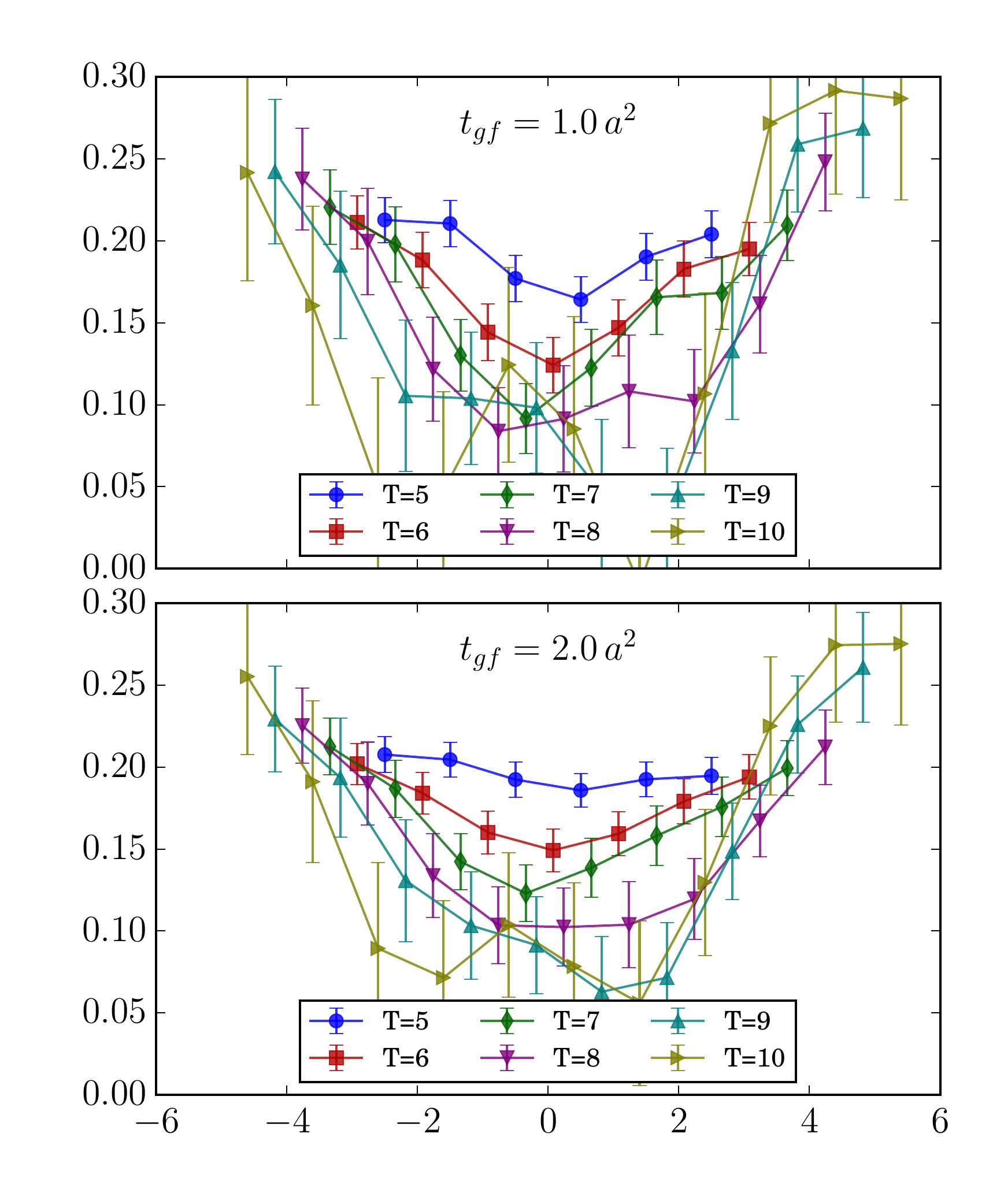} \ \ 
\includegraphics[width=0.49\textwidth]{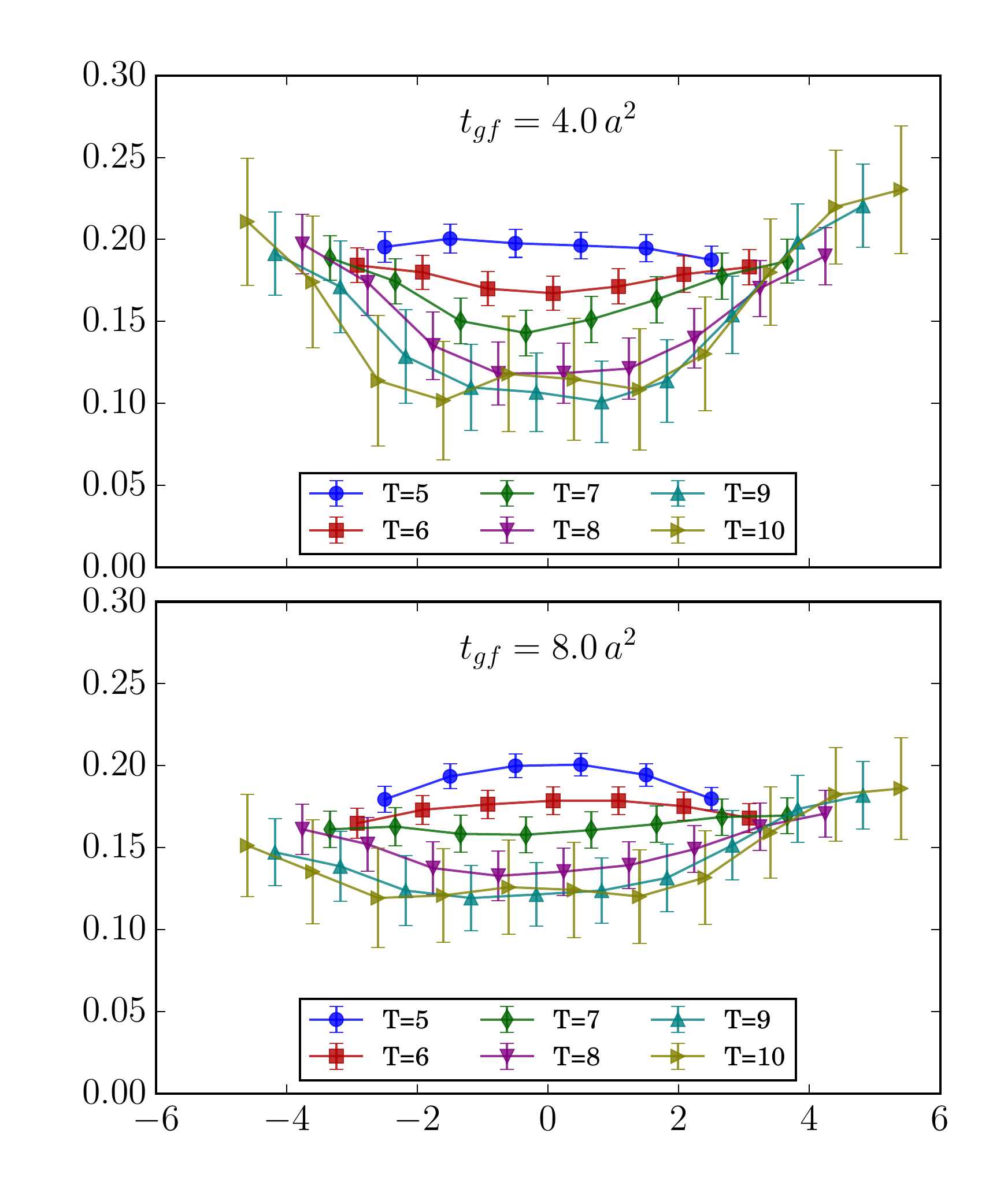}
\caption{
Gradient flow time dependence of $F_3(0)$ for each sink-source time separation $T$. 
The horizontal axis represents $t_Q-T/2$.
Results are obtained on a $24^3\times 64$ lattice with $m_\pi=330$ MeV 
and $|n|=2$ (the electric field quanta).
\label{fig:wft}}
\end{figure}

\begin{figure}[htb]
\centering
\includegraphics[clip,width=0.4\textwidth]{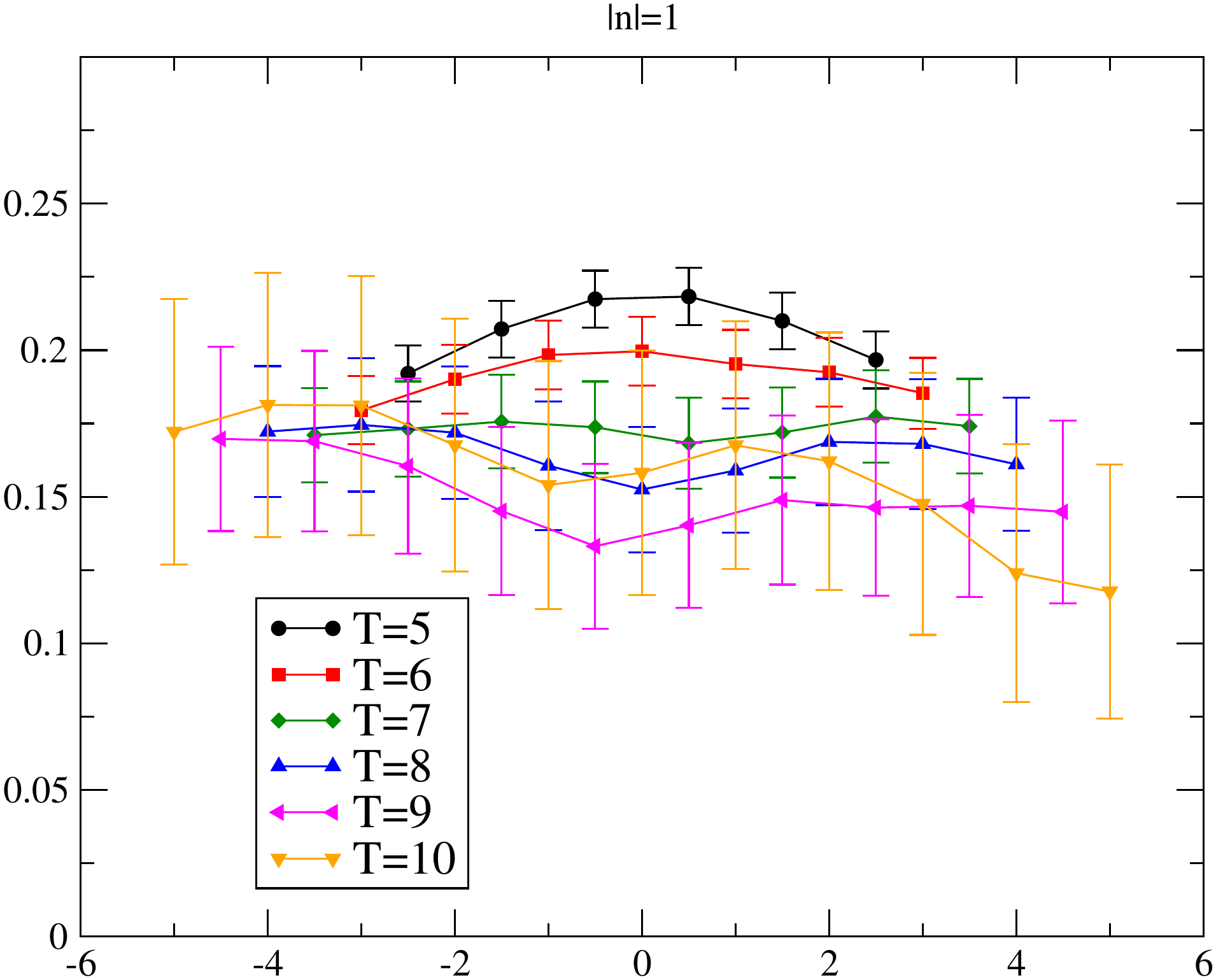}
\quad \quad 
\includegraphics[clip,width=0.4\textwidth]{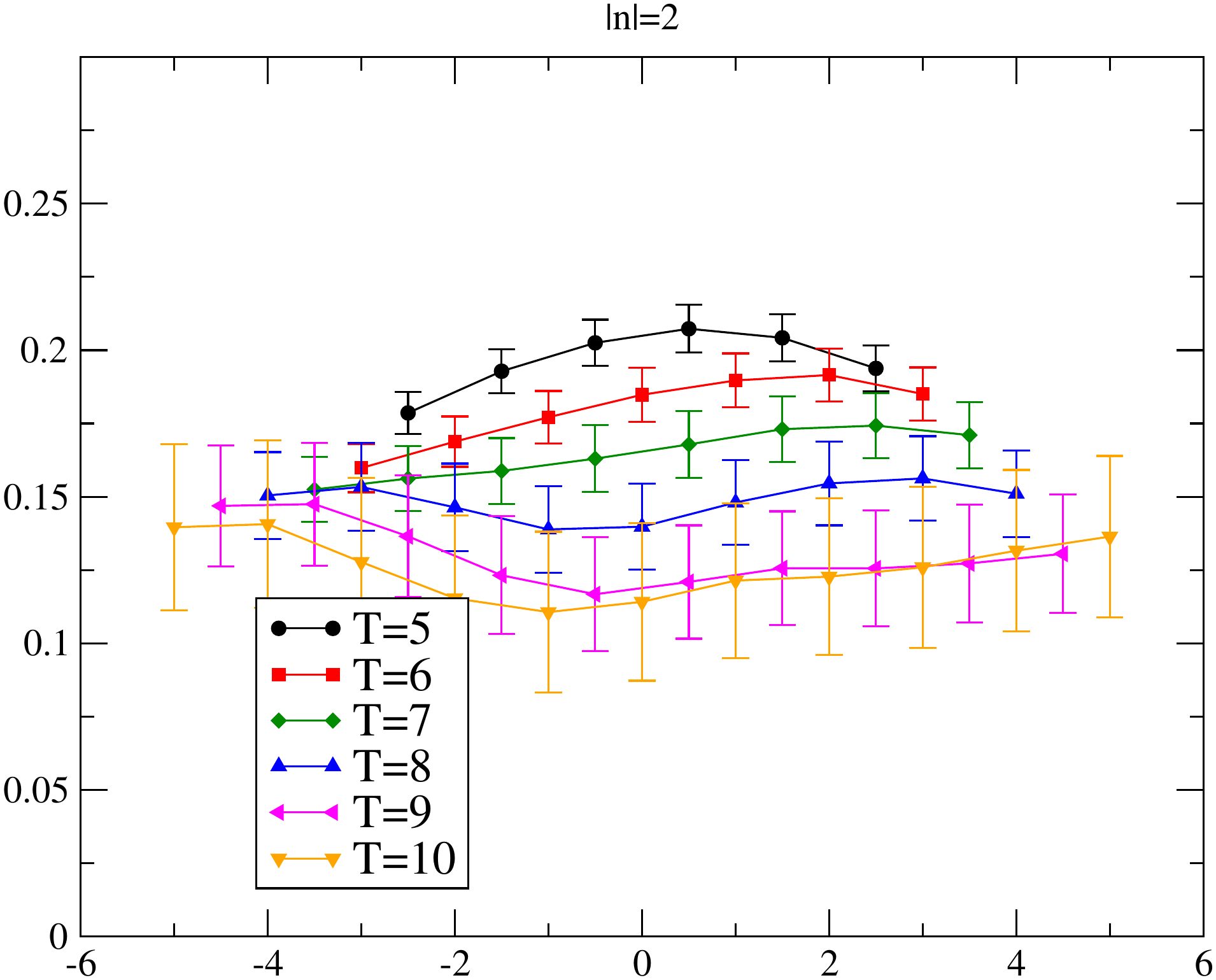}
\caption{
Electric field dependence of $F_3(0)$ for  $|n|=1$ and $2$ at a fixed gradient flow time. 
The horizontal axis represents $t_Q-T/2$.
Results are obtained on a $24^3\times 64$ lattice with $m_\pi=430$ MeV.
\label{fig:efd}}
\end{figure}

We show our preliminary results on the $\theta$-induced nEDM from the matrix element approach.
Since we compute the matrix element with the electric fields along $z$-direction in both positive and negative, 
we have four results for each component of $\pm E_z$ and projections $T_{S_{z\pm}}$ (see Fig.~\ref{fig:component}). 
Obviously these data are correlated with each other and 
differences between spin up (or positive $E_z$) and spin down (or negative $E_z$) reduce the error.
Fig.~\ref{fig:wft} shows the gradient flow time $t_{gf}$ dependence of $F_3(0)$
for each sink-source time separation $t=T$. 
As expected, the signal becomes better as increasing the flow time, 
and the result for $F_3(0)$ becomes stable at $t_{gf} \geq 4$ and $T \geq 8$.
We also study a possible higher order correction in $\vec{E}$. 
Fig.~\ref{fig:efd} shows the comparison of two results for $F_3(0)$ with different electric field strength with $|n|=1$ and $2$, 
where both results $|n|=1$ and $2$ are consistent with each other for $T \geq 8$, 
This result indicates that $O(\vec{E}^2)$ corrections on $F_3(0)$ are small. 
From the plateau at $T=9$ with $t_{gf}=8$ we obtain $F_3(0) =0.12(3)$ for the neutron at $m_\pi=330$ MeV.
Our result seems to be consistent with the previous result obtained in the form factor method (see e.g., Fig.~\ref{fig:Yoon})
while our value is only available at $Q^2=0$ and not directly comparable with the previous results.

%%%%%%%%%%%%%%%%%%%%%%%%%%%%%%
%%%%%%%%%%%%%%%%%%%%%%%%%%%%%%

%%%%%%%%%%%%%%%%%%%%%%%%%%%%%%
%%%%%%%%%%%%%%%%%%%%%%%%%%%%%%

%%%%%%%%%%%%%%%%%%%%%%%%%%%%%%
%%%%%%%%%%%%%%%%%%%%%%%%%%%%%%

\section{Summary}

Lattice calculations of nEDM are important for 
interpreting CP-violation effects in EDM experiments 
and cosmological observations.
A number of groups are putting in the effort required for computing nEDM 
at the physical point using the form factor method. 
The nEDM induced by $\theta$-term has a large statistical noise in its correlation 
to the topological charge density, which is not suppressed at a large distance
due to its global nature.
To reduce the error several noise reduction techniques using the truncation of (space and) time region of the topological charge density 
have been employed. 
It is found that 
while the truncation reduces the error by a factor of 2 at a heavier pion mass, 
due to its poor convergence 
no significant improvement is observed at the physical pion mass.
There also are several systematic uncertainties in $F_3(Q^2)$.
The uncertainties of the discretization effect, $Q^2\to0$ extrapolation, and excited state contamination 
for $F_3(Q^2)$ have not been well understood, 
which seem to become significant near the physical point.
To control the systematic errors and to further improve the statistical signal,
we have proposed a new approach using the matrix element with background electric fields.
This method only requires a local topological charge density operator between sink and source positions, 
and thus can avoid the large topological noise at a large distance.
In addition, no $Q^2$ extrapolation is required
since the forward matrix element is directly obtained from the energy shift.
Our preliminary results have demonstrated that we can achieve statistically-significant signal at heavier pion masses
that are consistent with the previous results. 
This method can in principle be applied to any $\CPviol$ operators, 
in which the Weinberg's three-gluon operator especially is beneficial for this method, 
since there is no additional computation cost for the gluonic operator. 
We need further investigations at the physical point. 

\section*{Acknowledgments}
\vspace{-1mm}
The speaker would like to thank the organizers of Lattice2019 in Wuhan for the invitation.  
We would also like to thank Tanmoy Bhattacharya and Boram Yoon for sending us their materials and 
the fruitful discussions.
We are grateful for the gauge configurations provided by the RBC/UKQCD collaboration.
This research used resources of the Argonne Leadership Computing Facility, which is a DOE Office
of Science User Facility supported under Contract DE-AC02-06CH11357,
and Hokusai supercomputer of the RIKEN ACCC facility.
H.O. is supported in part by JSPS KAKENHI Grant Numbers 17K14309 and 18H03710.
S.S. is supported by the National Science Foundation under CAREER Award 
PHY-1847893.

\end{document}